\documentclass[10pt,prd,twocolumn,amssymb,amsmath,superscriptaddress,nofootinbib]{revtex4}

\usepackage{graphicx}
\usepackage{color}
\usepackage{url}
\usepackage[utf8]{inputenc}
\usepackage{gensymb}
\usepackage{newunicodechar}
\DeclareUnicodeCharacter{00B0}{\degree}
\newunicodechar{°}{\degree}

\newcommand{\command}[2]{}%{{\color{red} #1: #2}\\}

\begin{document}

\title{BOOST -- A Satellite Mission to Test Lorentz Invariance Using High-Performance Optical Frequency References}

\newcommand{\ZARM}{\affiliation{ZARM, University of Bremen, Am Fallturm 2, 28359 Bremen, Germany}}
\newcommand{\DLR}{\affiliation{DLR Institute for Space Systems, Robert-Hooke-Strasse 7, 28359 Bremen, Germany}}
\newcommand{\HUB}{\affiliation{Humboldt University Berlin, Institute of Physics, Newtonstrasse 15, 12489 Berlin, Germany}}
\newcommand{\ADS}{\affiliation{Airbus Defence and Space GmbH, 88039 Friedrichshafen, Germany}}
\newcommand{\LUH}{\affiliation{Leibniz University Hannover, Institute for Quantumoptics, Welfengarten 1, 30167 Hannover, Germany}}
\newcommand{\PUM}{\affiliation{Philipps-University Marburg, Hans-Meerwein-Strasse 4, 35032 Marburg, Germany}}

\author{Norman G\"urlebeck}
\ZARM
\DLR

\author{Lisa W\"orner}
\ZARM
\DLR

\author{Thilo Schuldt}
\DLR

%---------------------------------

\author{Klaus D\"oringshoff}
\HUB

\author{Konstantin Gaul}
\PUM

\author{Domenico Gerardi}
\ADS

\author{Arne Grenzebach}
\ZARM
\DLR

\author{Nandan Jha}
\LUH

\author{Evgeny Kovalchuk}
\HUB

\author{Andreas Resch}
\DLR

\author{Thijs Wendrich}
\LUH

%------------------------------

\author{Robert Berger}
\PUM

\author{Sven Herrmann}
\ZARM

\author{Ulrich Johann}
\ADS

\author{Markus Krutzik}
\HUB

\author{Achim Peters}
\HUB

\author{Ernst M. Rasel}
\LUH

\author{Claus Braxmaier}
\ZARM
\DLR


\begin{abstract}
	BOOST (BOOst Symmetry Test) is a proposed satellite mission to search for violations of Lorentz invariance by comparing two optical frequency references. One is based on a long-term stable optical resonator and the other on a hyperfine transition in molecular iodine. This mission will allow to determine several parameters of the standard model extension in the electron sector up to two orders of magnitude better than with the current best experiments. Here, we will give an overview of the mission, the science case and the payload.
\end{abstract}

\maketitle

\section{Introduction}

General relativity and quantum theory are experimentally justified theories describing nature. 
%However, presently they are not compatible with each other
One of the biggest challenges of contemporary theoretical physics is to formulate a theory capable of unifying both, see, e.g., \cite{Isham1997} and references therein. Such a theory of quantum gravity could additionally explain phenomena at the Planck scale. Amongst others, such a theory is expected to resolve the singularity residing in a black hole and provide insights into the very early history of our universe. Despite enormous efforts, a commonly accepted theory was not yet found although some candidates were suggested like loop quantum gravity, string theory, discrete approaches such as causal dynamical triangulations, and non-commutative geometry, see, e.g., \cite{Rovelli1998,Rovelli2008} and references therein. However, there is no experimental evidence of the quantum properties of spacetime yet, presumably due to the inaccessibility of the energy scale at which they become relevant. Thus, highly accurate experiments must be performed to detect the minuscule remnants of these effects in our currently available regimes.

Such alternative theories usually violate some of the fundamental assumptions of our current physical theories like the Lorentz invariance,
%, which is the property that the outcome of any non-gravitational experiment performed in a freely falling frame is independent of the velocity and of the orientation of that frame.
which is a basic building block of special relativity, where it holds globally. In general relativity, it is still satisfied locally. A detection of a violation of Lorentz invariance (LIV) or the determination of tighter upper bounds on such violations aids the future development of new theoretical frameworks. 

In order not to be limited to specific alternative theories, test theories were developed, which quantify and catalog LIVs, most notably the standard model extension (SME) \cite{Colladay1997,Colladay1998,Kostelecky2004} but also the Robertson-Mansouri-Sexl (RMS) theory \cite{Robertson1949,MansouriSexl1977,Mansouri1977,Mansouri1977b}. Whereas the first describes general Lorentz violations for each particle, the second deforms Lorentz transformations introducing, e.g., a frame dependence in the speed of light. The latter approach is kinematic, i.e., it describes the LIV but it does not provide alternative field equations from which these effects ensue.

The satellite mission BOOST (Boost Symmetry Test) plans to measure these LIVs with unprecedented sensitivity by comparing two highly stable frequency references aboard the satellite. One laser is stabilized to a length standard given by an optical resonator and the other to a hyperfine transition in molecular iodine \cite{Doeringshoff2017}. Both frequency standards will be compared over the course of the satellite orbits. Since the changes of the frequencies of those two references are affected differently by possible LIVs in these test theories, a beat measurement provides an estimates on the parameters involved, see Sec.\ \ref{sec:ScienceCase}.

Within BOOST several key technologies are used and developed further so that they can be transferred to fit future developments and space-based missions. The ultra-stable, highly precise frequency references developed for BOOST provide new and valuable options for probing the gravitational field of the Earth. For example, the Gravity Recovery and Climate Experiment - Follow on mission (GRACE-FO) determines the gravity-induced change in the distance between two satellites using a laser ranging instrument as technology demonstration, see \cite{Sheard2012}. Here, the laser source is frequency stabilized using an optical resonator developed by JPL and Ball Aerospace Inc., USA, see \cite{Thompson2011}. Similar concepts are considered for ESA's Next Generation Gravity Mission (NGGM). Another example is the gravitational wave detector LISA (Laser Interferometer Space Antenna), for which an optical resonator is the baseline laser frequency pre-stabilization \cite{LISA2017}. 
 
Global Navigation Satellite Systems (GNSS) such as GPS or Galileo require high-performance clocks onboard as main payload. Their timing signals are used for position determination on Earth. Thus, the frequency stability of these clocks is one limiting factor for the accuracy of the positioning. Whereas current GNSS use microwave clock technology like Cs or Rb clocks as well as H-maser, future systems can benefit from optical frequency references like the iodine reference developed for BOOST. 

Currently, different efforts are on the way for developing optical frequency references for space. Laser frequency stabilization to an optical resonator is investigated, e.g., within the ESA projects Optical Stabilizing Reference Cavity (OSRC) with NGGM as application, see \cite{OSRC}, as well as a clock laser for a Strontium lattice clock and High Stability Laser (HSL) again with an application for NGGM, see \cite{Nicklaus2014}. Further space developments are carried out by SODERN (France) \cite{Argence2012} and by JPL/Ball Aerospace with respect to the flight model development for GRACE-FO, see \cite{Thompson2011,Bachman2017}. The optical resonator for BOOST is based on the German Aerospace Agency (DLR) developments towards a long-term stable optical resonator setup on Elegant Breadboard (EBB) level and frequency stabilization to molecular iodine on EBB and Engineering Model (EM) level, see \cite{Sanjuan2015,Schuldt2016} and \cite{Doeringshoff2017,Schuldt2017} respectively. Within the JOKARUS project led by the Humboldt-University Berlin, an iodine-based system is currently integrated for a payload on a sounding rocket with a tentative launch in the beginning of 2018, see \cite{Schkolnik2017}. Note that the iodine frequency reference fulfill the frequency stability requirements for LISA and NGGM \cite{Doeringshoff2017,Schuldt2017}. 

Aside the novel techniques in the field of highly stable frequency references, advanced laser technologies will be developed for the project. Currently, only specific wavelengths are accessible using space-qualified sources. With the planned diode lasers, the accessible range of wavelengths is broadened while the lasers' budgets are reduced at the same time. Such lasers could be envisaged for a multitude of future missions as well as in Earth-bound laboratories. They are also developed in the scope of the atom interferometry sounding rocket mission MAIUS \cite{Schkolnik2016}.

The paper is organized as follows. In Sec.\ \ref{sec:MissionOverview}, we introduce the mission including the science case and the driving requirements. Sec.\ \ref{sec:Payload} gives an overview of the payload and provides instrument budgets. In  Sec.\ \ref{sec:PayloadSubsystems}, the payload subsystems are described, and the corresponding error sources are discussed together with the respective error mitigation strategies.

\section{Mission Overview}
\label{sec:MissionOverview}

The satellite mission BOOST searches for LIVs, in particular, regarding the dynamics of electrons and photons. It is currently considered by the DLR in the scope of the national large mission program. It is based on previous studies of the satellite mission proposals STAR, BOOST, and mSTAR, see \cite{Lipa2012,Milke2013,Gurlebeck2015,Schuldt2015}, respectively. The tentative schedule foresees a launch in 2025. 

Subsequently, we describe the science case and the derived mission requirements.

\subsection{Science Case}\label{sec:ScienceCase}
There are different test theories available to describe possible LIV. We describe here the expected results of BOOST in the RMS framework and the SME. A detailed calculation will be given elsewhere.

\subsubsection{Robertson-Mansouri-Sexl test theory}
In the RMS theory, a distortion of the Lorentz transformation between the preferred frame $\Sigma_{\rm PF}$, in which the speed of light $c_0$ is assumed to be isotropic, and the experiment's rest frame $\Sigma_{\rm S}$, which moves with the velocity $\vec v$ relative to $\Sigma_{\rm PF}$,  is introduced. The deviation from the ordinary Lorentz transformations depends to leading order in $\frac{\vec v}{c_0}$ on the three parameters $\alpha,~ \beta,~\gamma$ \cite{Robertson1949,Mansouri1977,MansouriSexl1977,Mansouri1977b}.
They measure a deviation from the time dilation, longitudinal length contraction, i.e.\ in the direction of $\vec v$, and transversal length contraction as they are predicted by special relativity, in which $\alpha=-\beta=-\frac12,~\gamma=0$. This leads to a speed of light $c$ that depends on the relative velocity $\vec v$ and orientation $\theta$ of the light path with respect to the preferred frame:
\begin{align}\label{eq:variation_c}
\begin{split}
		\frac{c(\theta,\vec v)}{c_0}=1&+ \left(\beta-\alpha-1\right)\frac{\vec
		v^{2}}{c^{2}_{0}}+ 
	\left(\frac 12 -\beta+\gamma\right)\frac{\vec v^{2}}{c^{2}_{0}}\sin^2\theta\\
	&+O\left(\left|\frac{\vec v}{c}\right|^3\right),
\end{split}
\end{align}
where we already assumed that $\vec v$ is small compared to the speed of light.
Note that this is the two-way speed of light, i.e., the light travels from an observer A to a mirror B and back to A. Thus, a convention on how to synchronize clocks as for a one-way measurement is not necessary.

Combinations of the RMS parameters are measured by the three classical experiments, see \cite{Michelson1887,Kennedy1932,Ives1938}: 1) The Michelson Morley experiment measures $\alpha_{\rm MM}=\frac 12 -\beta+\gamma$ using the variation of the orientation $\theta$, 2) the Kennedy-Thorndike experiment measures $\alpha_{\rm KT}=\beta-\alpha-1$ using the variation of the relative velocity $\vec v$, and 3) the Ives-Stillwell experiment measures the time dilation and hence, $\alpha_{\rm IS}=\alpha+\frac 12$, directly. The most stringent constraints are given in Tab.\ \ref{tab:bestcoeff}.

\begin{center}
\begin{table}[h!]
\caption{Current constraints for the experimental determination of the RMS coefficients.\label{tab:bestcoeff}}
\begin{tabular}{|c|c|l|}
\hline 
Parameter  & Current-best constraint & Reference\\
\hline 
&&\\[-1em] 
$\alpha_{\rm KT}$   & $(4.8 \pm 3.7)\cdot 10^{-8} $  & \cite{Tobar2010}  \\
$\alpha_{\rm MM}$  & $(4 \pm 8)\cdot 10^{-12}$  & \cite{Herrmann2009}\footnote{Recently, 
\cite{Nagel2015} gave the most precise constraints on orientation-dependent relative frequency changes $\Delta \nu/ \nu$ to $9.2 \pm 10.7 ~ 10^{-19}$, one order of magnitude better than in \cite{Herrmann2009}. Although in \cite{Nagel2015} the experiment was not evaluated in the RMS framework, this implies also approximately an order of magnitude of improvement in $\alpha_{\rm MM}$ since the experiment was carried out at the same location.}\\
$\alpha_{\rm IS}$  & $(-0.38 \pm 1.06)\cdot \,10^{-8}$ & \cite{Delva2017}\\
\hline
\end{tabular}
\end{table} 
\end{center}

Subsequently, we apply the RMS framework to the experiment planned with BOOST consisting of an optical resonator and an	 iodine clock. The dependence of both on the potential variation of the speed of light \eqref{eq:variation_c} will be evaluated and the science signal identified. 

The resonance frequency $\nu_{\rm OR}(\Sigma_{\rm S})$ of the optical resonator depends on its rest frame $\Sigma_{\rm S}$ and the value of the speed of light in that frame. The frequency of the laser stabilized on a hyperfine transition of the iodine molecule $\nu_{{\rm I}_2}$ on the other hand is determined to leading order by the non-relativistic Hamiltonian and, thus, is on this level of approximation independent of the speed of light and $\vec v$. In fact, at higher orders of approximation a dependence appears via the fine structure constant, which is, however, suppressed compared to the frequency variations in the optical resonator. It serves as an absolute reference in this context. Thus, a beat measurement between the two yields
\begin{align}\label{eq:RMS_delta_nu}
\frac{\delta\nu_{\rm RMS}}{\nu_{\rm OR}(\Sigma_{\rm PF})}=\frac{\nu_{\rm OR}(\Sigma_{\rm S})-\frac{1}2 \nu_{{\rm I}_2}}{\nu_{\rm OR}(\Sigma_{\rm PF})}=\frac{c(\theta,\vec v)}{c_0}-\frac{\nu_{{\rm I}_2}}{2\nu_{\rm OR}(\Sigma_{\rm PF})},
\end{align}	
where the latter term is a constant offset, which we will not measure. The frequency $\nu_{\rm OR}(\Sigma_{\rm PF})$ is the frequency of a hypothetical optical resonator at rest in $\Sigma_{\rm PF}$, which is used solely for scaling purposes. Note that the factor $1/2$ in front of the $\nu_{{\rm I}_2}$ is due to the fact that in the planned experiment the resonance frequency of the optical resonator is compared with a laser that is first frequency doubled and then stabilized on the hyperfine transition of the iodine as described below, cf.\ Fig.\ \ref{fig2}. 
Together with Eq.\ \eqref{eq:variation_c}, this beat signal varies with $\vec v$ over one orbit and allows to determine $\alpha_{\rm KT}$. In fact, at the frequencies detectable with BOOST $\vec v$ varies only due to change in the satellite's velocity, i.e., due to the changes of the direction of its velocity. The Michelson-Morley coefficient $\alpha_{\mathrm MM}$ will be obtained simultaneously at the same time. However, the sensitivity of BOOST will not suffice to improve on the best-known constraints for that parameter, cf.\ Tab.\ \ref{tab:bestcoeff}, and we will omit its discussion here for brevity. Nonetheless, $\alpha_{\rm MM}$ will be considered in the data analysis of the mission.

One drawback of the RMS theory is that it requires a preferred frame. Although this can be chosen in principle arbitrarily, it is usually taken to be the rest frame of the cosmic microwave background, where the radiation is to a high degree isotropic. Nonetheless, future observations with different physical settings might suggest another preferred frame. Even though the results obtained for one frame can be easily transformed into any other frame, this can involve also a loss of sensitivity. Here, we will choose an orbit, which is sensitive to any possible direction of the preferred frame. Moreover, the RMS theory does not describe new field equations, say, for the dynamics of photons.

\subsubsection{The standard model extension}\label{sec:SME}

Both issues of the RMS theory, the need for a preferred frame and the lack of new field equations, are overcome by the SME, which is nowadays the test theory of choice, see \cite{Colladay1997,Colladay1998,Kostelecky2004}. It extends the action of the standard model with terms violating the Lorentz invariance, thereby, describing modifications of the dynamics of all particles. To achieve comparability of the results of different experiments, the measurements are always referred to a natural Sun-centered celestial equatorial frame (SCF) ($X^1,\, X^2,\, X^3,\,T$), see e.g., \cite{Kostelecky2002,Bluhm2003}. The $X^3$-axis is aligned with Earth's axis of rotation and $X^1$ points to the vernal equinox on the celestial sphere. The axis $X^2$ is chosen such that this frame is right-handed. The center of the sun is chosen as the spatial origin of the SCF and the origin of the time axis is chosen as the vernal equinox in the year 2000. 

The frequency of the optical resonator depends on the dynamics of the photons and also on the electron sector of the SME, which, e.g, describes the modification of the length of the optical resonator. It was argued in \cite{Muller2003} that the latter effect is suppressed compared to the former. Thus, the optical resonator is essentially sensitive to the photon sector of the SME, which is summarized in the modified Maxwell equations, cf.\ \cite{Kostelecky2002}:
\begin{align}\label{eq:modifiedMaxwell}
\frac{\partial}{\partial x^{\mu_2}} F_{\mu_1}^{\mu_2}+(k_F)_{\mu_1\mu_2\mu_3\mu_4}\eta^{\mu_2\mu_5}\frac{\partial}{\partial x^{\mu_5}} F^{\mu_3\mu_4}=0,
\end{align}
where $F$ is the Faraday tensor, $\eta$ the Minkowski metric with the signature $(+,-,-,-)$, and the $\mu_i$ are Lorentz indices running from $0$ to $3$. They are raised and lowered with the Minkowski metric. The $x^{\mu_i}$ are the spacetime coordinates, where $x^0$ and $x^1,~x^2,~x^3$ denote the timelike and spacelike ones, respectively. Note that we used the Einstein summation convention. Whereas the first term in Eq.\ \eqref{eq:modifiedMaxwell} is the ordinary source-free Maxwell equation, the second term is the modification of the SME parametrized by the $k_F$ tensor, which will be measured by BOOST. We neglected already terms proportional to the vector $k_{AF}$, i.e., those modifications depending explicitly on the four potential $A_\mu$ as well, following \cite{Kostelecky2002}. 

On the other hand, the iodine frequency reference is sensitive to the electron sector governed by the standard Hamiltonian with the Lorentz invariance violating correction, which reads in the non-relativistic limit, cf.\ \cite{Kostelecky1999}:
\begin{align}\label{eq:Hamilton}
\begin{split}
 \delta H=& c^2 \left[-b_j+m_e d_{j0}-\frac 12 \epsilon_{jkl}\left(m_eg_{kl0}-H_{kl}\right) \right]\sigma^j\\
 &-\left[c_{jk}+\frac 12 c_{00} \delta_{jk}\right]\frac{p_jp_k}{m_e}+\\ 
 &\biggl(\frac{1}{2}\left[b_l + \frac{1}{2}m_e \epsilon_{lmn}g_{mn0} 
 \right]\delta_{jk}+\biggl[m_e \left(d_{0j}+d_{j0}\right)  \\
 &-\frac{1}{2}\left(
 b_j + m_e d_{j0} + \frac{1}{2} \epsilon_{jmn}(m_e g_{mn0} 
 +  H_{mn}) \right) \biggr]\delta_{kl}\\
 &-m_e\epsilon_{jlm}\left(g_{m0k} + g_{mk0}\right) 
 \biggr)\frac{p_j p_k}{m_e^2}\sigma^l,
\end{split}
\end{align}
where $m_e$ is the electron mass, $c$ the speed of light, $\frac{\hbar}{2}\sigma^j$ and $p_j$ the spin and momentum operator of the electron, respectively. $\epsilon_{ijk}$ is the totally antisymmetric Levi-Civita symbol and $\delta_{jk}$ the Kronecker symbol. The lower case Latin indices $j,\,k,\,l,\,m,\,n$ run over the three spatial directions $1,\,2,\,3$, whereas the index $0$ refers to the time-like one. Analogously to the Einstein summation convention, we sum over repeated Latin indices in formula \eqref{eq:Hamilton}. The Lorentz tensors $b_{\mu_1},~c_{\mu_1\mu_2},~d_{\mu_1\mu_2},~H_{\mu_1\mu_2},~g_{\mu_1\mu_2\mu_3}$ parametrize the LIV in the electron sector of the SME. 
Note that we neglected here already terms odd in the electron's momentum, which vanish in the molecule's rest frame, and constant terms, which do not contribute to a shift in the transition frequency.  

A detailed treatment of the iodine frequency reference in the formalism of the standard model extension, which will be presented elsewhere, shows that only the terms proportional to the diagonal terms of $c^{\rm L}_{\mu\nu}$ in the laboratory frame contribute to the overall shift of the frequency. This is due to the symmetries of the iodine molecule and the fact that all orientations of the iodine molecule contribute to the spectral line. The other terms either vanish or they yield a broadening of the line, which is not yet detectable. The transformation of these parameters $c^{\rm L}_{\mu\mu}$ to the tensor components $c^{\rm SCF}_{\mu\nu}$ in the sun-centered frame will, however, introduce also off-diagonal terms again. 

The combination of the expressions of the photon and the electron sector yields following the formalism of \cite{Bluhm2003} the beat signal of the form:
\begin{align}
\begin{split}
\frac{\delta\nu_{\rm SME}}{\nu}=&\sum\limits_{i=1}^3\sum\limits_{j=-3}^3\left[ S_{ij} \sin(i\omega_{\rm S}T+j\Omega_\oplus T) \right.\\ 
&\left.+
C_{ij} \cos(i\omega_{\rm S}T+j\Omega_\oplus T ) 
\right],
\label{eq:SME_delta_nu}
\end{split}
\end{align}
where $\omega_{\rm S}$ is the frequency corresponding to one satellite orbit and $\Omega_\oplus$ to one revolution of the Earth around the sun and $T$ the time in the SCF. The coefficients $S_{ij}$ and $C_{ij}$ depend on the coefficients of the LIV, the orbit and the orientation of the optical resonator as well as the modification of the transition energies in the iodine molecules. Although we derive these coefficients explicitly elsewhere, we give in the appendix two of them for illustration purposes. Note that $S_{1\pm3}=C_{1\pm 3}=S_{3\pm3}=C_{3\pm 3}=C_{30}=0$. This implies that there are in general $33$ fitting parameters to such a science signal or equivalently peaks in the power spectral density of the relative frequency. However, they will not all be independent and not all will be observable, i.e. they are already constrained by previous experiments below our noise limit, cf.\ \cite{Kostelecky2011}.\footnote{\label{fn:constraints}Note that some of these known constraints are also based on theoretical arguments like in the case of astrophysical birefringence, whereas BOOST would measure them directly. Nonetheless, we omit such constraints in the discussion below for brevity, cf.\ Tab.\ \ref{tab:scientificoutput}, and present them elsewhere.} 
Thus, comparing the $S_{ij}$ and $C_{ij}$ with the expected stability of the used references gives the estimates for the experimental outcome as will be discussed the next section.

\subsection{Science and Mission Requirements}

The science requirements that follow from the previous section are summarized subsequently. Of course the requirements on the orbit and the instrument are not entirely independent. Taking Eqs.\ \eqref{eq:RMS_delta_nu} and \eqref{eq:SME_delta_nu} into account, it is obvious that the variations take place at frequencies near the orbital frequency. Thus, the references have to perform well at this time scale.

Generally, an orbit with a low altitude is preferable for several reasons. First, the satellite's speed is higher for lower altitudes. This gives together with the change of the direction of the velocity of the satellite over one orbit higher velocity variations, which will be beneficial for both test theories. Second, since during one complete orbit one estimate of the different constraints of the test theories can be generated, the statistics is improved with a lower altitude implying more orbits per day if a similar relative frequency stability is assumed at orbit time. 
Both effects are also the main reasons why this experiment is more sensitive to LIVs if carried out on a satellite rather than on ground: For a low-Earth orbit (LEO) this amounts to an improvement by roughly two orders of magnitude if the same experiment is carried out for the same period in the laboratory or aboard a satellite.

Moreover, shorter orbital periods entail a less restrictive requirement on the stability of the frequency references, which is especially important for the optical resonator. If the altitude becomes too low, however, the atmospheric drag will shorten either the lifetime of the mission or increase its complexity by the need to reposition the spacecraft. Thus, a low-Earth orbit below the inner van Allen belt ($1000$\,km), where the sensitivity varies only by a few percent with the altitude, is preferable. 

In order to be able to resolve the different frequencies in Eq.\ \eqref{eq:SME_delta_nu} in a Fourier analysis of the science data, the mission should be in science mode for at least one year. Assuming a duty cycle of about $50\%$ a mission lifetime of $2$ years is required. To allow later an appropriate data analysis, like in \cite{Herrmann2009} for example, the satellite should operate ten full orbits in science mode without disturbances. Nonetheless, we will assume here a continuous science mode of the satellite for one year in the science case evaluation consistently with the level of approximations done subsequently.

We want the experiment to be sensitive to all possible directions of the preferred frame in the RMS theory. In the SME, this is equivalent to requiring to be sensitive to all spatial components of the tensors measuring Lorentz violation like $c_{\mu\nu}$. This leads to an orbit where the orbital plane sweeps out the entire space in the course of one year, which is guaranteed with a sun-synchronous orbit. This reduces also eclipses for the satellite and relaxes the requirements on the thermal control system and power management of the satellite. 

The analysis of different orbit options indicated that a $6$am dawn-dusk sun-synchronous orbit at $650$\,km altitude is a good compromise satisfying the aforementioned constraints, guaranteeing the necessary sensitivity level for the science signal, and the need to reduce the impact by drag effects. Moreover, the remaining eclipse time is reduced even further and with this choice the satellite can de-orbit freely in $25$ years as required for the space debris mitigation. The ground visibility is acceptable, too. 

The orientation of the optical resonator should be chosen such that the orientation of the optical paths change over one orbit, which enhances the time variability of the science signal in the SME evaluation. Hence, one optical path should be pointing in the direction of the relative velocity of the satellite with respect to the Earth and the other one parallel to its relative acceleration, i.e., nadir pointing. Assuming that the optical resonator is mounted rigidly to the spacecraft, this implies an attitude for the satellite, where the angles between the satellite axes and the optical paths are fixed.\footnote{In \cite{Bluhm2003}, this is called XVV mode.}
Note that this is not required for measuring the Kennedy-Thorndike coefficient in the RMS theory.

With this orbit, the scientific output can be predicted, cf.\ Tab.\ \ref{tab:scientificoutput} as follows. Requiring a relative frequency stability of the references of $1\cdot 10^{-15}$ at orbit time and assuming white noise in the relevant frequency regime, an expected power spectral density (PSD) can be derived for a one-year mission that is continuously in science mode. This PSD is then compared to Eqs.\ \eqref{eq:RMS_delta_nu} and \eqref{eq:SME_delta_nu}, which determines constraints for the coefficients $S_{ij}$, $C_{ij}$ and $\alpha_{\rm KT}$. Afterwards, these constraints can be converted to constraints on the SME parameters with straight forward algebra.

For these estimates, we neglect terms which are already constrained below our noise level. Hence, only those, which improve the current best estimates by up to two orders of magnitude, see \cite{Kostelecky2017}, are shown here, cf.\ Footnote \ref{fn:constraints}. The instrument requirements derived from this science requirement are discussed in the next section.

\begin{center}
	\begin{table}[h!]
		\caption{Expected constraints on LIV by the proposed mission BOOST after one year of observation.  %The $\kappa_{e-}^{ij}$ are linear combinations of the components of the tensor $k_F$, which parametrizes the Lorentz violation in the photon sector, see, e.g., \cite{Kostelecky2002} for their definition. 
			%In particular, $\kappa_{e-,{\rm SCF}}^{33}=-\frac{4}{3} (k_F)^{\rm SCF}_{1212}-\frac{1}{3}
			%(k_F)^{\rm SCF}_{1313}+(k_F)^{\rm SCF}_{1010}-\frac{1}{3}(k_F)^{\rm SCF}_{2323}+(k_F)^{\rm SCF}_{2020}$.  
			\label{tab:scientificoutput}}
		\begin{tabular}{|c|}
			\hline 
			Constraints\footnote{Note that the precision of the constraints of the SME parameters is limited, e.g., by the precision of the estimates of the expectation value of the perturbation of the Hamiltonian in Eq.\ \eqref{eq:Hamilton}.} \\
			\hline 
			\\[-1em]
			$\left|c^{\rm SCF}_{10}+c^{\rm SCF}_{01}\right|\leq 3\cdot 10^{-13}$\\
			$\left|c^{\rm SCF}_{30}+c^{\rm SCF}_{03}\right|\leq 3\cdot 10^{-13}$\\
			$\left|c^{\rm SCF}_{12}+c^{\rm SCF}_{21}\right|\leq 4\cdot 10^{-17}$\\
			$\left|c^{\rm SCF}_{13}+c^{\rm SCF}_{31}\right|\leq 2\cdot 10^{-17}$\\
			$\left|c^{\rm SCF}_{23}+c^{\rm SCF}_{32}\right|\leq 3\cdot 10^{-17}$\\
%			$\left|c^{\rm SCF}_{11}-c^{\rm SCF}_{22}\right|\leq 4\cdot 10^{-17}$\\
%			$|c^{\rm SCF}_{33} - 0.99 (c^{\rm SCF}_{11} + c^{\rm SCF}_{22} ) + 4.9 \cdot 10^{-6} (c^{\rm SCF}_{20}+c^{\rm SCF}_{02}) +$\\
%			$2.3\cdot 10^{-5} (c^{\rm SCF}_{30}+c^{S\rm CF}_{03}) +0.82 \kappa_{e-,{\rm SCF}}^{33}|\leq 1.1 \cdot 10^{-17}$\\ 
			$\left|c^{\rm SCF}_{11}+c^{\rm SCF}_{22}-2c^{\rm SCF}_{33}\right|\leq 4\cdot 10^{-17}$\\
			$\left|\alpha_{\rm KT}\right|\leq 7.5\cdot 10^{-10}$\,\footnote{The value for $\alpha_{\rm KT}$ is referring to the rest frame of the cosmic microwave background as preferred frame. Preferred frames in directions orthogonal to this one yield analogous results provided they move at the same speed with respect to us, which is just a scaling for comparability.} \\
			\hline
		\end{tabular}
	\end{table} 
\end{center}

\section{Payload Overview}
\label{sec:Payload}

\begin{figure}%
\includegraphics[width=\columnwidth]{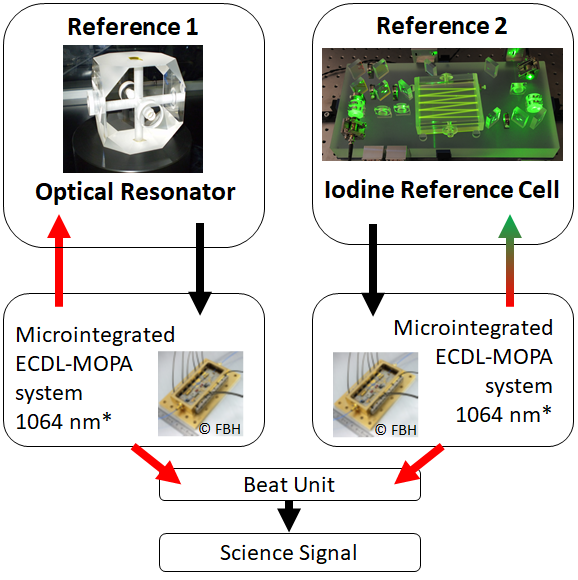}
\caption{Schematic overview of the measurement principle. An optical resonator and an iodine spectroscopy unit are employed to stabilize their respective lasers developed by the Ferdinand-Braun-Institute (FBH). The resulting stabilized frequencies are then compared in the beat measurement. The time variation of the beat signal yields the science signal. ($^{*}$ cf.\ \cite{Schkolnik2017}.)}%
\label{fig2}%
\end{figure}

To measure the small deviation in the photon and electron propagation, the scientific payload consists of two optical frequency references -- an optical resonator and an iodine spectroscopy unit. Both frequency references shall operate with a relative stability of $10^{-15}$ at orbit time, i.e.\ approximately $90$\,min. 
A sketch of the measurement principle can be seen in Fig.\,\ref{fig2}.

In this section, an overview of the flight hardware, including the thermal and redundancy concept as well as the budgets, will be given. The following section, Sec.\,\ref{sec:PayloadSubsystems}, then describes the payload subsystems including the possible error sources and the respective mitigation strategies in more detail.

\subsection{Thermal and Redundancy Concept}

\begin{figure*}%
\centering
\includegraphics[width=0.8\textwidth]{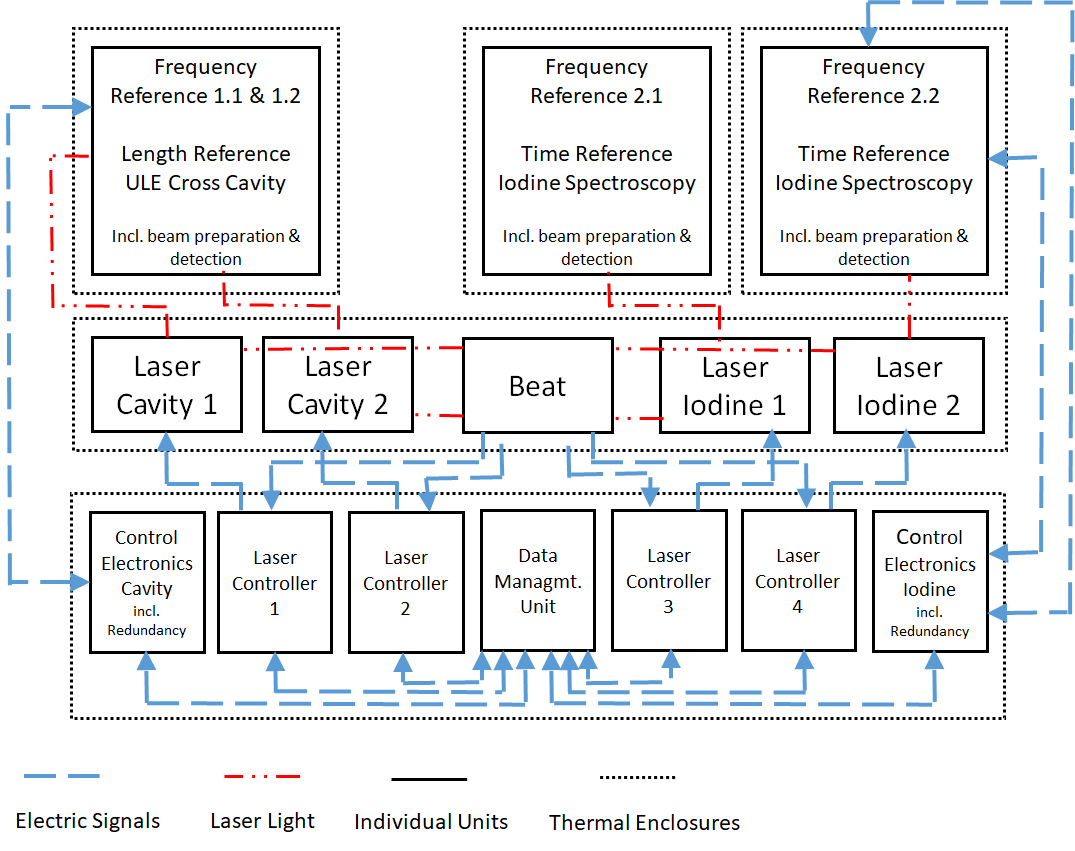}%
\caption{Schematic overview of the payload. The beat unit as well as the data management unit are internally redundant.}%
\label{fig1}%
\end{figure*}

A schematic of the payload is given in Figure\,\ref{fig1}. Along this scheme, we will explain the thermal and redundancy concept. 

The \textit{thermal stability} of the payload is a major factor in the performance of the instrument subsystems. While the mass and power budgets could be reduced using one large compartment, housing the entire payload, the easiness of implementation into the satellite bus and the mitigation of potential thermal noises induced by one of the other systems favor individual thermal stabilization of the subsystems.

As it can be seen in Fig.\,\ref{fig1}, five thermally stabilized compartments are chosen as baseline for the payload's design with individual compartments for the optical resonator, the iodine spectroscopy units, the laser system, and the control electronics, respectively. To avoid the impact of thermal fluctuations, the beam preparation and detection stages are implemented into the same housing as the payload subsystem, i.e. the optical resonator and the iodine spectroscopy unit. 
%Noise introduced by the optical fibers connecting the laser system to the payload subsystems will be mitigated by active fiber stabilization. 

For \textit{redundancy}, the two frequency references are doubled. The redundancy concept is sketched in Fig.\,\ref{fig1}. In case of the optical resonator, a spacer with two crossed light paths is chosen implementing the redundancy of the optics in one ultra-low expansion glass (ULE) block. Both accessible optical paths are equipped with a beam preparation and detection stage. They are housed in one thermally stabilized box and they are used to stabilize two individual lasers. In contrast, two complete iodine spectroscopy units in separate boxes are included in the payload. Each system is associated with one dedicated laser. All four lasers are connected to the beat unit. This allows to compare each of the iodine spectroscopy units to each of the optical paths of the resonator. Nonetheless, to reduce the power and ease the requirements on the batteries during eclipse times, cold redundancy is chosen as baseline for the payload. 

\subsection{Mass and Power Budgets}

The resulting overall budgets for the payload are summarized in Table\,\ref{tab3}. All of the values given in this table include $20\%$ component level margin. An additional $20\%$ system level margin is added to the total budget of the payload. The mass and the power reflect the cold redundancy concept described above.

\begin{table}%
\caption{The payload budgets including $20\%$ component level margin and an additional $20\%$ system level margin on the total budget.\label{tab3}}
\begin{tabular}{|l|c|c|c|}
\hline
Item & \# Units & Mass & Power\\ 
\hline
Optical Resonator & 1 & 57\,kg & 11\,W\\ 
\hline
Iodine Spectroscopy & 2 & 14\,kg & 12\,W\\
\hline
Laser and Beat & 1 & 15\,kg & 15\,W\\
\hline
Electronics & 1 & 44\,kg & 186\,W\\
\hline
Harness & 1 & 26\,kg & 0\,W\\
\hline
&&&\\
\hline
\textbf{Total incl. $20 \%$ margin} & & \textbf{204\,kg} & \textbf{269\,W}\\
\hline
\end{tabular}
\end{table}

\section{Payload Subsystems} 
\label{sec:PayloadSubsystems}

\subsection{Optical Resonator Unit} 

Optical resonators are employed to stabilize lasers using the Pound-Drever-Hall scheme\,\cite{Drever1983}. Within BOOST, a cubic optical resonator based on the NPL design \cite{Webster2011} is chosen, cf.\ Fig.\ \ref{fig2}. The spacer of the optical resonator will be made out of ULE and the mirrors of fused silica to reduce the thermal noise and the sensitivity to external thermal fluctuation. The spacer is mounted at four points with tetrahedral symmetry as in \cite{Webster2011} to reduce the vibration sensitivity. We will choose the curvature radii of the mirrors to be $1$\,m and $\infty$ respectively.  We deviate from the NPL design by choosing a longer path length of $8.7$\,cm in order to reduce the thermal noise floor. The mass and volume limitations of a space mission constraint the length, although a longer baseline would reduce the thermal noise floor further. Additionally, for the specific length and curvature radii of the mirrors, the higher TEM modes are sufficiently separated from another to ensure that the modulation frequency of the Pound-Drever-Hall side bands can be chosen such that they do not overlap with those modes. The cube is designed in a way that two optical paths can be operated at any given time. 

Current state-of-the-art optical resonators achieve a frequency stability in the order of several parts in $10^{-17}$ on time scales from one tenth of a second up to several seconds \cite{Matei2017}. However, optical cavities that have been designed specifically for space applications and high robustness demonstrate a frequency stability of $10^{-15}$ at $1$\,s \cite{Swierad2016, Abgrall2015}. For BOOST, we require on the other hand stabilities of $10^{-15}$ at $90$\,min, which requires additional developments. Subsequently, the major limitations and mitigation strategies to achieve this frequency stability are discussed.

\textit{External thermal fluctuations} have a high impact on the long-term stability of the resonator if they are not attenuated since any length variation due to thermal expansion translates directly into a frequency variation. To counteract the occurring thermal fluctuations, two measures are taken. First, the spacer is made from ULE, which has generally a low coefficient of thermal expansion (CTE) and in particular a zero crossing of the CTE. The optical resonator is then operated near this zero crossing temperature of the CTE. Second, a five-fold thermal shield is mounted around the resonator for a passive attenuation of external temperature fluctuations. Five aluminum shields with a thickness of $3$\,mm each are calculated to attenuate the temperature fluctuations by a factor of $10^{5}$\ at $90$ min, see \cite{Sanjuan2015}. Additionally, the outer shields' temperature is actively stabilized to $\pm 1$\,mK at a temperature that is in the $10$\,mK range of the CTE zero crossing. The thermal shields are separated by Ti-spacers and the holes for the optical access covered with BK7 glass to reduce the temperature fluctuations to a minimum. The materials are chosen based on their thermal conductivity and transparency to the chosen wavelength. A detailed description of the chosen materials including the impact of the properties and design can be found in Ref.\,\cite{Sanjuan2015}. A linear frequency drift due to isothermal relaxation of the ULE will be removed from the signal. 

Each of the optical resonators' components contribute to the \textit{thermal Brownian noise limit}. Taking the size and the materials of the mirror substrate and coatings as well as of the ULE spacer into account, the resulting thermal noise floor is estimated to $3.9 \cdot 10^{-16}$, cf.\ \cite{Kessler2012, Kessler2012_1}. 
Indeed, this is the highest contribution to the total noise.

Additionally, frequency fluctuations are introduced via \textit{intensity fluctuations} of the in-coupled light onto the mirror substrate. These fluctuations are typically in the order of $100-200$\,Hz/$\mu$W\, see \cite{Haefner2015}. Assuming laser intensity fluctuations in the order of $0.5$\,nW, the frequency fluctuations in the optical resonator are no higher than $3.5\cdot 10^{-16}$ at orbit time.

The \textit{residual amplitude modulation} (RAM) is another source for frequency fluctuations on the long time scale required by the experiment. The RAM is therefore stabilized actively. Considering a finesse of $4\cdot 10^5$ for the optical resonator and a RAM stabilization of $2\cdot10^{-5}$ at $90$\,min, the limit to the achievable frequency stability is $3\cdot 10^{-16}$, see \cite{Zhang2014}.

Furthermore, the refractive index and thereby the optical path length is influenced by \textit{pressure density fluctuations} along the optical paths. To avoid these, the resonator is placed inside a vacuum chamber. The frequency fluctuation caused by pressure fluctuations of $10^{-9}$\,mbar at a base pressure of $10^{-8}$\,mbar is below $2.7\cdot 10^{-16}$\, cf.\ \cite{Birch1994}.

Other error sources for the optical resonator are \textit{gravity induced distortions} in the optical resonator, \textit{residual accelerations} caused by vibrations, rotation of the satellite and orbital drag, \textit{de-modulator phase instabilities}, and \textit{electronic noises}. All of these effects contribute in the range of $10^{-17}$ or below to the frequency noise of the optical resonator.

\begin{table}%
\caption{Error Budgets for the Optical Resonator.\label{tab1}}
\begin{tabular}{|l|c|c|}
\hline
\multicolumn{1}{|l|}{Noise sources} & \multicolumn{1}{|c|}{ $\frac{\delta\nu}{\nu}\cdot 10^{16}$} & \multicolumn{1}{|c|}{Ref.}\\[0.1cm]
\hline
Thermal fluctuations & $1$&\cite{Sanjuan2015}\\
Thermal Brownian noise & $3.9$&\cite{Kessler2012}\\
Intensity fluctuations & $3.5$&\cite{Haefner2015}\\
Residual amplitude modulation & $3$&\cite{Zhang2014}\\
Pressure fluctuations & $2.7$&\cite{Birch1994}\\
Gravity gradient & $0.1$&\cite{Laemmerzahl2004}\\
De-Modulator phase instability & $2$ & \cite{Nevsky2001, Hong2004}\\
Vibrations & $0.25$ & \cite{Webster2011}\\
\hline
\textbf{Total} & \textbf{$7.0$}&\\
\hline
\end{tabular}
\end{table}

The error budgets for the optical resonator are combined in Table\,\ref{tab1} assuming that the individual contributions are independent from one another. %The achievable relative frequency stability is given by the root Allan deviation $\sigma_{\frac{\delta \nu}{\nu}}$. 
The aforementioned frequency noises limit the performance of the optical resonator below the required relative frequency stability of $\frac{\delta\nu}{\nu}$ of $10^{-15}$ at orbit time. In the worst case, if all noise sources except the Brownian noise would couple fully, say, via temperature fluctuations, they would sum up to $2\cdot 10^{-15}$. 

\subsection{Iodine Spectroscopy Unit} 

In the iodine spectroscopy unit a hyperfine transition of di-atomic iodine at $532$\,nm is used to stabilize the laser via Doppler-free saturation spectroscopy\,\cite{Shirley1982}. For these frequency references, a performance at $10^{-15}$ stability level on long time scales has been established\,\cite{Doeringshoff2017, Schuldt2017}. 
In further efforts, compact units for space based applications have been developed\,\cite{Schkolnik2017, Schuldt2017}. The molecular iodine will be held in a compact multi-pass gas cell with an interaction length of approximately $90$\,cm. The spectroscopy setup is realized using a glass baseplate where the optical components are integrated by adhesive bonding. Subsequently, we discuss the major limitations to the stability at orbit time.
 
Among other factors, the achievable frequency stability of the iodine spectroscopy depends on the \textit{line width} of the transition at $532$\,nm, which is in the order of $200-300$\,kHz, see \cite{Cheng2002}. Given the accessibility of this wavelength using lasers at $1064$\,nm, operating the iodine spectroscopy at $532$\,nm is the practical choice. The hyperfine transition at $508$\,nm has a natural line-width of $50-100$\,kHz, see \cite{Cheng2002}. Thus, the performance of the spectroscopy could be enhanced addressing this narrower line of the hyperfine spectrum. However, the currently available laser modules have a better performance at $532$\,nm, which is, thus, chosen as baseline. 

The performance of the iodine frequency reference is limited by the \textit{gas pressure} inside the gas cell to $-2.2$\,kHz/Pa, see \cite{Schuldt2017}. Since the gas pressure is regulated via a cold finger, this translates to a fluctuation in its temperature of $-300$\,Hz/K, see \cite{Schuldt2017}.
With the required stability of the cold finger of $1$\,mK, this results in a stability of $5\cdot 10^{-16}$ at orbit time. 

Variations in the \textit{laser power} induce a shift in the molecular resonance frequency. Typically, this results in a frequency fluctuation of $300$\,Hz/mW, see \cite{Nevsky2001, Hong2004}. Assuming $10$\,mW of laser power, cf.\ \cite{Schuldt2017}, and fractional intensity fluctuations of $1\cdot 10^{-4}$, the impact of the resulting frequency calculations can be estimated as $3.5\cdot 10^{-16}$ at orbit time.

The \textit{modulation transfer spectroscopy} (MTS) signal slope was measured in the laboratory setup at Humboldt University Berlin. The corresponding coefficient is in the range of $200$\,Hz/mV. 
Following the requirement that the electronic offset fluctuations shall not be higher than $1$\,$\mu$V, the resulting frequency fluctuation is $3.5 \cdot 10^{-16}$.

\textit{Residual amplitude modulation} is another source of frequency fluctuations in iodine systems\,\cite{Burck2009}. If the RAM contribution can be limited to $1 \cdot10^{-7}$ at orbit time, the resulting frequency fluctuations will be limited to $4.2 \cdot10^{-16}$ at orbit time, see \cite{Jaatinen1998, Jaatinen2009}. This is a rather stringent requirement but it may be close to realization considering recent performance levels of iodine frequency standards reaching below the $3 \cdot10^{-15}$ level, cf.\ \cite{Schuldt2017}.

The stability of the \textit{angle between the pump and probe beam }introduces frequency fluctuations.
With a decoupling of $25$\,mrad and a frequency shift of $2$\,kHz/mrad, a frequency fluctuation of $3.5\cdot 10^{-16}$ can be expected using adhesive bonding, see \cite{Ressel2010}. 

Other effects, such as \textit{phase modulation index fluctuations}, \textit{de-modulator phase instabilities}, and \textit{external magnetic field fluctuations} further contribute to the limitation of the performance of the iodine spectroscopy. The contributions for the most important error sources are displayed in Table\,\ref{tab2}. 

\begin{table}%
\caption{Error Budgets for the Iodine Frequency Reference\label{tab2}}
\begin{tabular}{|l|c|c|}
\hline
\multicolumn{1}{|l|}{Noise sources} & \multicolumn{1}{|c|}{ $\frac{\delta\nu}{\nu}\cdot 10^{16}$} & \multicolumn{1}{|c|}{Ref.}\\[0.1cm]
\hline
Pressure fluctuations & $5 $ & \cite{Schuldt2017}\\
Light power fluctuations & $3.5 $ & \cite{Nevsky2001, Hong2004}\\
Servo electronic offsets & $3.5 $ & \footnote{As measured with the engineering model setup \cite{Schuldt2017} at the Humboldt University Berlin.}\\
Residual amplitude modulation & $4.2 $ & \cite{Jaatinen1998, Jaatinen2009}\\
Beam pointing instability & $3.5 $ & \cite{Ressel2010}\\
Phase modulation fluctuations & $3$ & \cite{Cordiale2000}\\
De-Modulator phase instability & $2$ & \cite{Nevsky2001, Hong2004}\\
Magnetic field fluctuations & $1$ & \cite{Goncharov1995, Goncharov1996}\\
\hline
\textbf{Total} & \textbf{$9.7 $} &\\
\hline
\end{tabular}
\end{table}

\subsection{Laser and Beat Unit}

The laser sources for BOOST are based on a micro-integrated diode laser technology platform developed at the Ferdinand-Braun-Institute (FBH) in a joint lab activity with the Humboldt-University Berlin.
This platform provides compact, robust and energy-efficient semiconductor laser modules with the advantage of broad wavelength accessibility \cite{Wicht2017}. Other wavelengths (e.g.\ $508$\,nm) might be of interest for addressing hyperfine spectra near the B-state dissociation limit of molecular iodine. These diode laser modules operate in experiments at the Bremen drop tower to study ultra-cold atomic gases \cite{Schiemangk2015}, and have been used in several sounding rocket missions to realize optical frequency references \cite{Lezius2016,Dinkelaker2017} as well as the first Bose-Einstein condensate in space \cite{Schkolnik2016,MAIUS}. Currently, a compact iodine  frequency reference is prepared for launch in April 2018 aboard the TEXUS 54 sounding rocket as an important qualification step towards space application \cite{Schuldt2017}.

A part of the laser output, which is stabilized with the optical resonator or the iodine spectroscopy unit, is then routed to the beat unit. By observing the beat note, differences between the frequencies can be observed.
Depending on the analysis the observed deviation is then linked to the respective parameters in the above discussed test theories. 
The quality of the beat measurement, thus, impacts the generated science signal.

The stability of the beat measurement is governed by the stability of the implemented radio frequency (RF) source.
With the targeted relative frequency stability of $10^{-15}$ at orbit time and a free spectral range of the optical resonator of about $2$\,GHz, a stability of $1\cdot10^{-11}$ at orbit time for the RF source is required including margin. This can be established by employing the Chip Scale Atomic Clock (CSAC) as RF source. In consequence, an addition to the achievable frequency stability of $10^{-16}$ caused by the accuracy of the beat has to be taken into account. 

Another reduction of the frequency stability is due to the individual housing of the payload subsystems. In this design, the lasers are housed in an enclosure separated from the optical resonator and the iodine frequency references, respectively. Thus, the fibers, connecting the laser system to the according frequency reference, are exposed to thermal fluctuations. The satellite bus shall be stabilized to $\pm 5$\,K. With a fiber length of $0.5$\,m, this introduces a frequency instability of $10^{-16}$ at orbit time\,\cite{Musha1982, Toyoda1983}. %The current design foresees an active fiber stabilization, which eliminates this additional noise.  
\\

\section{Summary}

We discussed the satellite mission BOOST, which will test the Lorentz Invariance in space. It is a candidate mission in the Large Mission framework of the DLR. We showed that this mission would improve our current best measurements of the parameters of the SME, in particular in the electron sector, by one to two orders of magnitude. Moreover, we demonstrated the feasibility of such an experiment in terms of performance of the individual frequency references, their beat and availability of components. The details of the experiment as well as mission parameters like the satellite platform and the possible launch options will be discussed elsewhere.

\section{Acknowledgments}

This work is supported by the German Space Agency DLR with funds provided by the Federal Ministry for Economic Affairs and Energy under grant number 50 OO 1604, 50 OO 1605, 50 QT 1401, 50 QT 1201 and 50 QT 1102.

\appendix*

\section{Science Signal}\label{sec:Appendix}

The science signal Eq.\ \eqref{eq:SME_delta_nu} contains $33$ fitting parameters $S_{ij}$ and $C_{ij}$. We give here two examples for illustration purposes:
\begin{widetext}
\begin{align}
\begin{split}
C_{10}=&\frac{R_S  \sin(\zeta)}{32 \nu_0  \pi} \bigg[\left(\Delta \frac{p^2_z}{2m_e}  (6 \omega_S - \Omega_\oplus \cos(\zeta)) + 
\Delta \frac{p^2_x}{2 m_e} (14 \omega_S + \Omega_\oplus \cos(\zeta))\right)(c^{\rm SCF}_{30} + c^{\rm SCF}_{03}) - \\
& 16  \pi (2 \omega_S + \Omega_\oplus \cos(\zeta)) \kappa_{o+,{\rm SCF}}^{12}\bigg]\\
C_{32}=&\frac{R_S \Omega_\oplus}{64 \nu_0} (5 \sin(\zeta) + 4 \sin(2\zeta) + \sin(3\zeta)) \kappa_{o-,{\rm SCF}}^{1 2}
\end{split}
\end{align}
\end{widetext}
The appearing constants have the following meaning, cf.\ \cite{Bluhm2003}:
$\zeta$ is the angle between the Earth's rotation axis, i.e., the $X^3$-axis in the SCF, cf.\ Sec.\ \ref{sec:SME}, and the normal of the satellite's orbit. For the considered orbit, this is $97\degree$.
%$\eta\approx 23\degree$ denotes the angle between Earth's rotation axis and its orbital plane. 
$R_S\approx 3.5 \cdot 10^{13}$\,eV${}^{-1}$ 
%and $R_\oplus\approx 7.6\cdot 10^{17}$\,eV${}^{-1}$ 
is the radius of the satellite's circular orbit.
% and the radius of the Earth's orbit around the sun, respectively
Here as with the rest of the appendix, natural units are employed as it is common in the SME. The angle $\alpha$ is the azimuthal angle between the satellite plane and the orbital plane of the Earth measured from the $X^1$-axis of the SCF frame. For a sun-synchronous orbit like we consider here, it behaves like $\alpha=\alpha_0 +  \Omega_{\oplus} t$. This was already employed to derive Eq.\ \eqref{eq:SME_delta_nu}. $\alpha_0$ is a constant that is determined by the choice of the origin of the time coordinate and the launch date of the satellite and is chosen to vanish here for convenience. Not that we also assume here an optical resonator, where one optical axis is parallel to the relative velocity of the satellite with respect to the Earth and the other is nadir pointing.

$\Delta \frac{p^2_x}{2m_e}\approx -1\cdot 10^1$\,eV and $\Delta \frac{p^2_z}{2m_e}\approx 3\cdot 10^1$\,eV are abbreviations for rough estimates\footnote{Note that the precision of the final results in Tab.\ \ref{tab:scientificoutput} is limited by these estimates to one significant digit.} of the difference of the expectation values of the operators of the kinetic energy in the respective directions for the two states $X^1\Sigma^+_g$ and $B^3\Pi_{0+u}$ involved in the absorption. These estimates correspond to the molecule's rest frame, which is oriented such that the $x^3-$axis is along the molecules axis. $\nu_0=18.56$\,eV is the frequency of the unperturbed laser.

The $\kappa_{o-,{\rm SCF}}^{ij}$ are linear combinations of $\left(k_{F}\right)^{\rm SCF}_{\mu_1\mu_2\mu_3\mu_4}$, see, e.g.\ \cite{Kostelecky2002}. They are already well constraint by astrophysical tests, cf.\ \cite{Kostelecky2017}, so that they appear in the fitting parameters $S_{ij}$ and $C_{ij}$ only below our noise limit, which is the reason why we omitted them in Tab.\ \ref{tab:scientificoutput} for brevity, cf.\ Footnote \ref{fn:constraints}. Under this assumption, $C_{10}$ yields the second constraint in Tab.\ \ref{tab:scientificoutput}. Interestingly, $C_{32}$ is just affected by the SME modifications of the photon sector. 

%\bibliographystyle{h-physrev}
%\bibliography{biblio}

\end{document}